\documentclass[namedreferences]{solarphysics}

\usepackage[hyperref,optionalrh]{spr-sola-addons} 
\usepackage{graphicx}        
\usepackage{color}           
\usepackage{breakurl}        

\chardef\us=`\_

\graphicspath{{./}{figures/}}

\begin{document}

\begin{article}
\begin{opening}

\title{Polarimetric Studies of a Fast Coronal Mass Ejection\\ }

\author[addressref={aff1,aff2},corref,email={marilena.mierla@oma.be}]{\inits{M.}\fnm{Marilena}~\lnm{Mierla}\orcid{0000-0003-4105-7364}}
\author[addressref=aff3,email={inhester@mps.mpg.de}]{\inits{B.}\fnm{Bernd}~\lnm{Inhester}\orcid{0000-0002-1066-0411}}
\author[addressref={aff1,aff4},email={andrei.zhukov@oma.be}]{\inits{A.N.}\fnm{Andrei N.}~\lnm{Zhukov}\orcid{0000-0002-2542-9810}}
\author[addressref={aff1},email={sergei.shestov@oma.be}]{\inits{S.V.}\fnm{Sergei V.}~\lnm{Shestov}\orcid{0000-0003-2849-6282}}
\author[addressref=aff5,email={alessandro.bemporad@inaf.it}]{\inits{A.}\fnm{Alessandro}~\lnm{Bemporad}\orcid{0000-0001-5796-5653}}
\author[addressref=aff6,email={philippe.lamy@latmos.ipsl.fr}]{\inits{P.}\fnm{Philippe}~\lnm{Lamy}\orcid{0000-0002-2104-2782}}
\author[addressref=aff7,email={koutchmy@iap.fr}]{\inits{S.}\fnm{Serge}~\lnm{Koutchmy}}

\address[id=aff1]{Solar-Terrestrial Centre of Excellence - SIDC, Royal Observatory of Belgium, Avenue Circulaire 3, Brussels 1180, Belgium}
\address[id=aff2]{Institute of Geodynamics of the Romanian Academy, Bucharest, Romania}
\address[id=aff3]{Max-Planck-Institute for Solar System Research G\"{o}ettingen, Germany}
\address[id=aff4]{Skobeltsyn Institute of Nuclear Physics, Moscow State University, 119992 Moscow, Russia}
\address[id=aff5]{INAF-Turin Astrophysical Observatory, Turin, Italy}
\address[id=aff6]{Laboratoire Atmospheres, Milieux et Observations Spatiales, CNRS \& UVSQ, Guyancourt, France}
\address[id=aff7]{Institut d'Astrophysique de Paris - CNRS and Sorbonne University, France}

\runningauthor{M. Mierla et al.}
\runningtitle{Polarimetric Studies of a Fast CME}

\begin{abstract}
In this work we performed a polarimetric study of a fast and wide coronal mass ejection (CME) observed on 12 July 2012 by the COR1 and COR2 instruments onboard \textit{Solar TErrestrial RElations Observatory} (STEREO) mission. The CME source region was an X1.4 flare located at approximately S15W01 on the solar disk as observed from the Earth's perspective. The position of the CME as derived from the 3D Graduated Cylindrical Shell (GCS) reconstruction method was at around S18W00 at 2.5 solar radii and S07W00 at 5.7 solar radii, meaning that the CME was deflected towards the Equator while propagating outward in the corona. The projected speed of the leading edge of the CME also evolved from around 200\,km s$^{-1}$ in the lower corona to around 1000\,km s$^{-1}$ in the COR2 field of view.
The degree of polarisation of the CME is around 65\,\% but it can go as high as 80\,\% in some CME regions.
The CME showed deviation of the polarisation angle from the tangential in the range of 10$^\circ$\,--\,15$^\circ$ (or more). Our analysis showed that this is mostly due to the fact that the sequence of three polarised images from where the polarised parameters are derived is not taken simultaneously, but at a difference of few seconds in time. In this interval of time, the CME is moving by at least two pixels in the FOV of the instruments and this displacement results in uncertainties in the polarisation parameters (degree of polarisation, polarisation angle, etc.). 
We propose some steps forward to improve the derivation of the polarisation.	
This study is important for analysing the future data from instruments with polarisation capabilities. 
\end{abstract}
\end{opening}

\section{Introduction}
     \label{S-Introduction} 

The white-light emission of the solar K-corona originates in Thomson-scattering of photospheric light by free electrons in the corona. Detailed description of the Thomson-scattering theory can be found in various articles \citep[see, e.g.,][]{Minnaert1930, Baumbach1938, vandeHulst1950, Shklovskii1965, Billings1966, Inhester2015}.
 
It is known that the degree of polarisation of Thomson-scattered light by coronal electrons is a sensitive function of the scattering angle between the incident light direction and the direction towards the observer \citep{Billings1966}. This effect allows us to estimate an effective distance of the coronal scattering volume off the plane of the sky (POS) and it has been employed to obtain 3D position of coronal mass ejections (CMEs) from coronagraph images obtained with different polariser orientations \citep[see, e.g.,][]{Moran2004, Dere2005, Vourlidas2006, Mierla2009}.

In a recent study, \citet{Floyd2019}, taking advantage of the much improved determination of the polarisation of the corona \citep[see, e.g.,][]{Lamy2014} from the LASCO-C2 images \citep{Brueckner1995}, applied the polarised-ratio technique to a set of 15 CMEs observed in the time interval from 1998 to 2014. The authors showed that despite the unfavorable conditions affecting the LASCO-C2 polarisation measurements, the polarimetric reconstruction works quite well and is able to produce meaningful 3D CME morphology and direction of propagation. Even in the case of the very fast CME of 18 July 2002 (speed around 2000\,km s$^{-1}$), its localization could be obtained from its bulk part.

Because the observed polarised brightness is much less sensitive to stray light, it has been also widely used to estimate the electron density of the average corona and of the total mass of CMEs \citep[see, e.g.,][]{Quemerais2002, Feng2015, Decraemer2019}.
As the estimation of the CME mass and propagation direction has important implications for space weather forecasts it is important to have accurate measurements of the polarisation parameters quantities.

The accurate measurements of the polarisation parameters can be checked by the tangential orientation of the polarisation direction. There are limits to this check, both experimental (mirrors may rotate the polarisation \citep{Floyd2019}, atmospheric effects to ground measurements \citep[see][]{Koutchmy1999}, limited precision, etc.), and theoretical (non-random orientation of dust, relativistic electrons \citep{Molodensky1973, Kishonkov1975}, magnetically sensitive resonant line scattering \citep[see, e.g.,][]{Mierla2011, Qu2013, Heinzel2020}. 
In this paper we present the observation of a systematic deviation of the polarisation angle. This deviation occurs near the propagation front of a fast CME. This lead us to the conclusion that we have to account for yet another effect which may lead to rotated polarisation angles. This effect has so far not been considered in the literature. We discuss this effect quantitatively and compare it with possible alternative reasons in the discussion section.
The article is structured as follows: In Section~\ref{S-obs} we introduce the data, and in Sections~\ref{S-data} and \ref{S-dataanalysis} we describe how they are processed and analysed. Discussions and conclusions are presented in Section~\ref{S-Discussions} and \ref{S-Conclusions}, respectively.

\section{Observations} 
\label{S-obs} 
In this work we use COR1 and COR2 images taken on 12 July 2012. 
COR1 is the coronagraph with the innermost field of view of the \textit{Sun Earth Connection Coronal and Heliospheric Investigation} (SECCHI) instrument
suite \citep{Howard2008} onboard the twin \textit{Solar TErrestrial RElations Observatory spacecraft} \citep[STEREO: see][]{Kaiser2008}. Each of the STEREO/COR1 telescopes has a field of view from 1.4 to 4 solar radii (R$_{\odot}$) and observes in a white-light waveband 22.5\,nm wide centred at the H$\alpha$ line at 656\,nm \citep{Thompson2008}. 

Like the COR1, the SECCHI outer field of view coronagraph, COR2, observes the coronal signal in visible light from 2 to 15\,R$_{\odot}$, with a passband from 650 to 750\,nm. 
Both COR1 and COR2 coronagraphs take polarised images at three different polarisation angles at 0$^\circ$, 120$^\circ$, and 240$^\circ$. 

The COR1 images have a dimension of $512\times512$ pixels and a pixel size of 15$^{''}$. The exposure time is one second and the image cadence five minutes. Each set of three polarised images ($I_{0}$, $I_{120}$, $I_{240}$) is taken in approximately 18 seconds (COR1-A) and 24 seconds (COR1-B), with nine seconds between the frames in the case of COR1-A data, and 12 seconds between the frames for COR1-B.

The COR2 images have a dimension of $2048\times2048$ pixels and a pixel size of 15$^{''}$. The exposure time is six seconds and the image cadence one hour. Each set of three polarised images ($I_{0}$, $I_{120}$, $I_{240}$) is taken in approximately 60 seconds with 30 seconds between the frames.

On 12 July 2012 a fast and wide CME (associated with an X1.4 flare) was observed at the East limb on COR-A images, at the West limb in COR-B images and coming from the central disk in LASCO images.  
The approximate location of the CME source region (the X1.4 flare) was S15W01 as observed from the Earth's perspective (see, \textsf{e.g.}, \url{https://cdaw.gsfc.nasa.gov/CME\_list/halo/halo.html} and \url{https://solarmonitor.org/index.php?date=20120712&region=11520}).

The CME was observed first time by COR1 at around 16:00 UT, by COR2 at around 16:40 UT and by LASCO-C2 at around 16:50 UT.
COR1 and COR2 coronagraphs observed a structured, three-part CME consisting of a bright circular leading edge, followed by a dark cavity and a bright compact core \citep[see, e.g.,][]{Illing1986}, while LASCO-C2 and -C3 coronagraphs observed a full-halo CME with apparent angular width of $360^\circ$. 
A diffuse shock \citep[see, e.g.,][]{Vourlidas2003} was also observed ahead of the leading edge.

\section{Data Processing}
\label{S-data}

Data processing consists of three steps:

\begin{enumerate}[i)]
\item Pre-process the data using the IDL SolarSoft \textsf{secchi\_prep.pro} routine.
\item Extract the K-corona signal by subtraction base-difference image.
\item Extract the polarised parameters -- polarised brightness, total brightness, degree of polarisation, deviation angle from tangential -- from three K-corona images with different polarisation $I_{0}$, $I_{120}$, $I_{240}$.\\
\end{enumerate}


\subsection{Data Pre-processing}
\label{S-preprocessing}

We run \textsf{secchi\_prep} on each polarised COR1 and COR2 image. By default, \textsf{secchi\_prep} applies the following calibration to COR images: division by exposure duration; correction for onboard image processing; subtraction of the CCD bias; multiplication by the calibration factor to convert the image values from digital number (DN) to mean solar brightness (MSB); vignetting and flat-field correction; optical distortion correction (COR2 only).

The COR1 and COR2 images were brought to the same dimension: $1024\times1024$ pixels with the pixel size of 7.5$^{''}$ for COR1 and 29.4$^{''}$ for COR2. 

We divided each image by the calibration factor $6.780 \times 10^{-11}$ for COR1 and $1.350 \times 10^{-12}$ for COR2 to have units of DN s$^{-1}$. We preferred DN s$^{-1}$ as it offers a more straightforward way to display the data. The choice of the units does not influence our results. 


\subsection{Extracting K-corona Signal} 
\label{S-kcorona}
  
In general, one raw white-light coronagraph image contains signal from the solar corona (K-corona: electron or continuum corona; F-corona: Fraunhofer or dust corona; E-corona: emission lines like H$\alpha$, Fe\,{\sc x}, Fe\,{\sc xiv}, Ca\,{\sc xv}, He\,{\sc I} D3 line etc.), from the cosmic rays, planets, comets, stars, as well as from the instrument related contribution: bias, dark current, stray-light, etc. Some of this signal is polarised, and some is unpolarised.

K-corona signal is due to the scattering of the sunlight (photospheric light) by the free electrons of the corona also known as the Thomson-scattered light. The F-corona is the scattering of the light on the dust particles around the Sun (contains Fraunhofer absorption lines). The E-corona is formed from the emission of highly ionised atoms of the solar corona. One more component, the thermal (T-corona) is produced by thermal emission of dust particles heated by the Sun \citep{Shopov2008}.

The K-corona contains features like streamers, CMEs, outflows, inflows, etc.
In the ideal case we would like to get rid of all the unwanted contribution and keep the signal only from the K-corona, or in our case only from the CME Thomson-scattered light. 
In order to get rid of the unwanted signal, one can employ different procedures: (i) subtraction of the monthly minimum background from a raw white-light coronagraph image; (ii) subtraction of a pre-event image from a raw white-light coronagraph image (the so-called base-difference image technique); (iii) subtraction of a hourly-minimum background from a raw white-light coronagraph image. 

These procedures are described in detail in Appendix~\ref{S-bkg}.

For this study we used base-difference images by subtracting the frame before the CME was observed, at 15:50 UT in COR1 and at 16:09 UT in the COR2 field of view (FOV). Three different pre-event images are used, one each for the three polarisers. Further on if not stated otherwise we will always be using base-difference images.

The influence of different backgrounds on our results is discussed in Appendix~\ref{S-bkginfl}. 

In Figure~\ref{F-3polariz} it is shown an example of one set of three polarised COR1-A images taken on 12 July 2012, 16:35 UT.

\begin{figure}	
	 \includegraphics[width=0.9\textwidth]{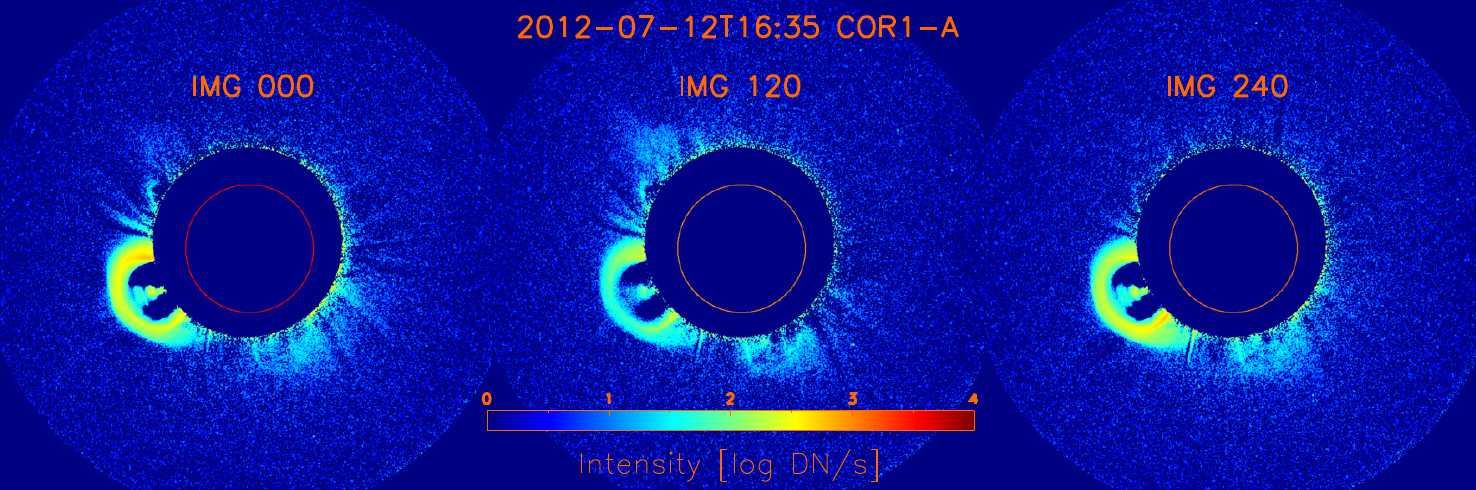}
	\caption{COR1 images taken on 12 July 2012, with orientation of the polariser of $0^\circ$ taken at 16:35:00 UT (\textit{left panel}), $120^\circ$ at 16:35:09 UT (\textit{middle panel}), and $240^\circ$ at 16:35:18 UT (\textit{right panel}). Images are displayed in logarithmic scale. Base-difference images were used. The radius of the occulter is $1.47$\,R$_{\odot}$ and the circle inside the occulter outlines the solar disk.} \label{F-3polariz}
\end{figure}


\subsection{Extraction of the Polarisation Parameters} 
\label{S-extrpolparam}

From COR1 and COR2 polarised images processed as explained above we derived polarised quantities like: total brightness (\textit{B} -- Equation \ref{E-tb}), polarised brightness (\textit{pB} -- Equation \ref{E-pb}),
the degree of polarisation (\textit{p} -- Equation \ref{E-p}), and the polarisation angle orientation (\textit{$\alpha$} -- Equation \ref{E-ang}) of the K-corona signal.  
The deviation angle from tangential (\textit{$\Delta \alpha$}) is calculated as the difference between the tangent to the solar limb and the polarisation angle. Positive values of the deviation angle correspond to the clockwise rotation and the negative values to the counter-clockwise rotation. 

\begin{equation}\label{E-tb}
  B = \frac{2}{3} \,(I_{0} + I_{120} + I_{240})
\end{equation}

\begin{equation}\label{E-pb}
  pB = \frac{4}{3} \,\sqrt{(I_{0} + I_{120} + I_{240})^2 - 3\,(I_{0}I_{120} + I_{0}I_{240} + I_{120}I_{240})}
\end{equation}


\begin{equation}\label{E-p}
  p = \frac{pB}{B}
\end{equation}

\begin{equation}\label{E-ang}
 \alpha = \frac{1}{2} \,\arctan{\frac{\sqrt{3}\,(I_{240} - I_{120})}{2I_{0} - I_{120} - I_{240}}}
\end{equation}

These formulas correspond to frequently used ones that utilize Stokes parameters \textit{I}, \textit{Q}, and \textit{U}. Here we use explicit formulas to simplify analysis presented in Section~\ref{S-devang}. In Equation \ref{E-tb} and Equation \ref{E-pb} we used the directions of $0^{\circ}$, $120^{\circ}$, and $240^{\circ}$ of the polarisers, instead of the usual direction of $0^{\circ}$, $60^{\circ}$, and $120^{\circ}$ because these are the angles at which the COR1 polarisers are reported to be oriented, \citep[see, e.g.,][]{Thompson2008}. Note that $60^{\circ}$, and $240^{\circ}$ polariser orientations are identical.

The $I_{0}$, $I_{120}$, and $I_{240}$ represent the images recorded by the COR1 instrument at three polarisation angles, from which a pre-event image was subtracted as mentioned in the previous section. In this way, the three polarised images $I_{0}$, $I_{120}$, and $I_{240}$ contain largely only the component of the CME K-corona.

The polarisation angles for COR2 images were corrected with the values provided in the SolarSoft routine \textsf{cor1\_fitpol.pro} \citep{Thompson2015}. This was necessary because the COR2 polarisers do not have a specific alignment.

\section{Data Analysis and Interpretation}
\label{S-dataanalysis}

\subsection{CME Polarisation Parameters} 
  \label{S-polparam}
  
\begin{figure}	
    \centering
    \includegraphics[width=0.85\textwidth]{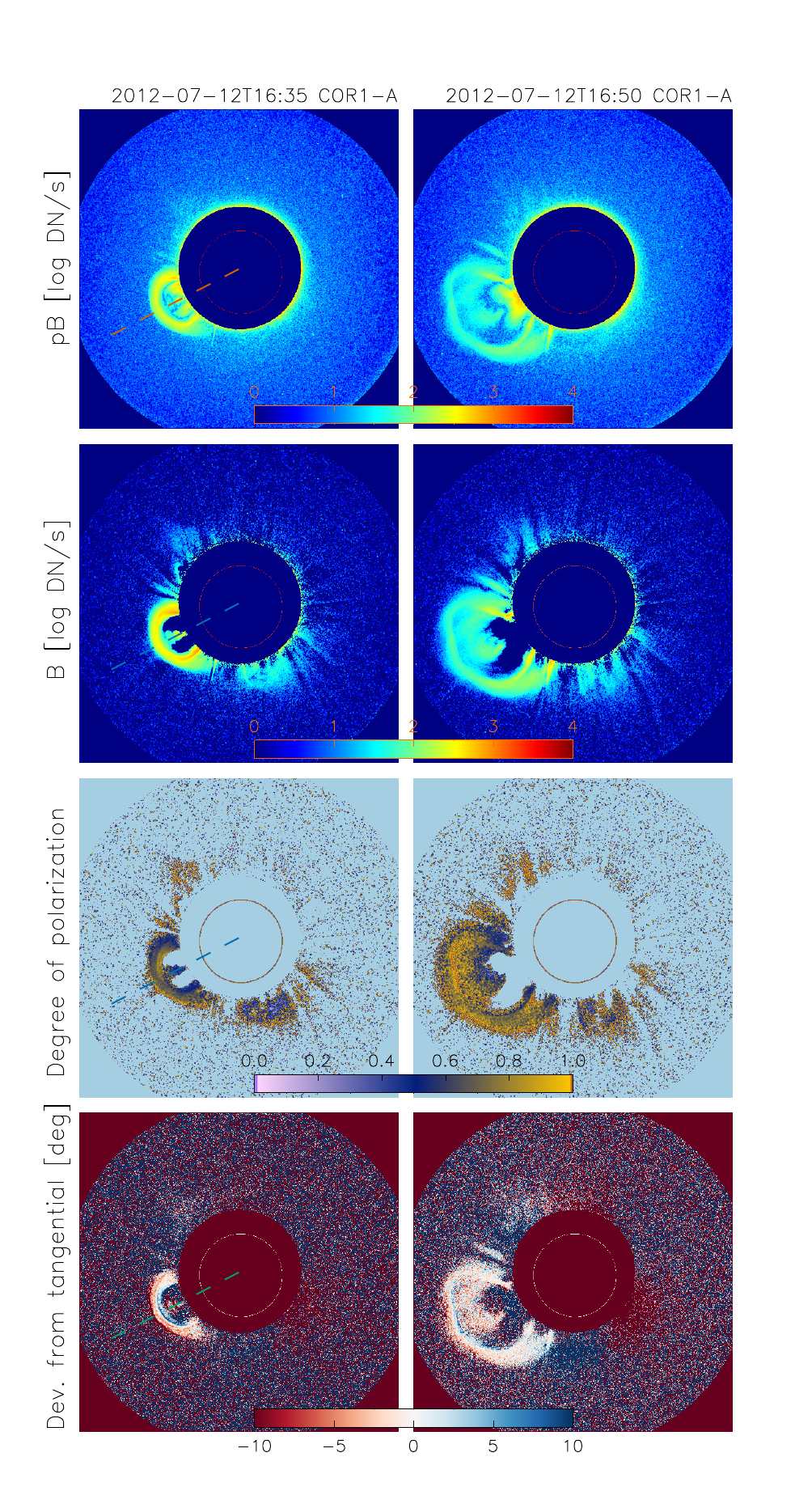}
    \caption{COR1-A polarisation parameters derived from polarised images on 12 July 2012, 16:35 UT (\textit{left panels}) and 16:50 UT (\textit{right panels}) showing: \textit{pB}-images (\textit{upper panels}), \textit{B}-images (\textit{middle-upper panels}), degree of polarisation (\textit{middle-lower panels}) and deviation from tangential angle (\textit{lower panels}). \textit{pB} and \textit{B} images are displayed in logarithmic scale. The parameters are derived from base-difference images. The circle inside the occulter outlines the solar disk.} \label{F-polarizparam}
\end{figure}

In Figure~\ref{F-polarizparam} we show the extracted polarisation parameters from COR1-A images taken at 16:35 UT (left column) and at 16:50 UT (right column), as following: polarised brightness $pB$ (upper panels), total brightness $B$ (middle-upper panels), degree of polarisation $p$ (middle-lower panels), and deviation angle from tangential (lower panels). 

\begin{figure}	
    \centering
     \includegraphics[width=0.9\textwidth]{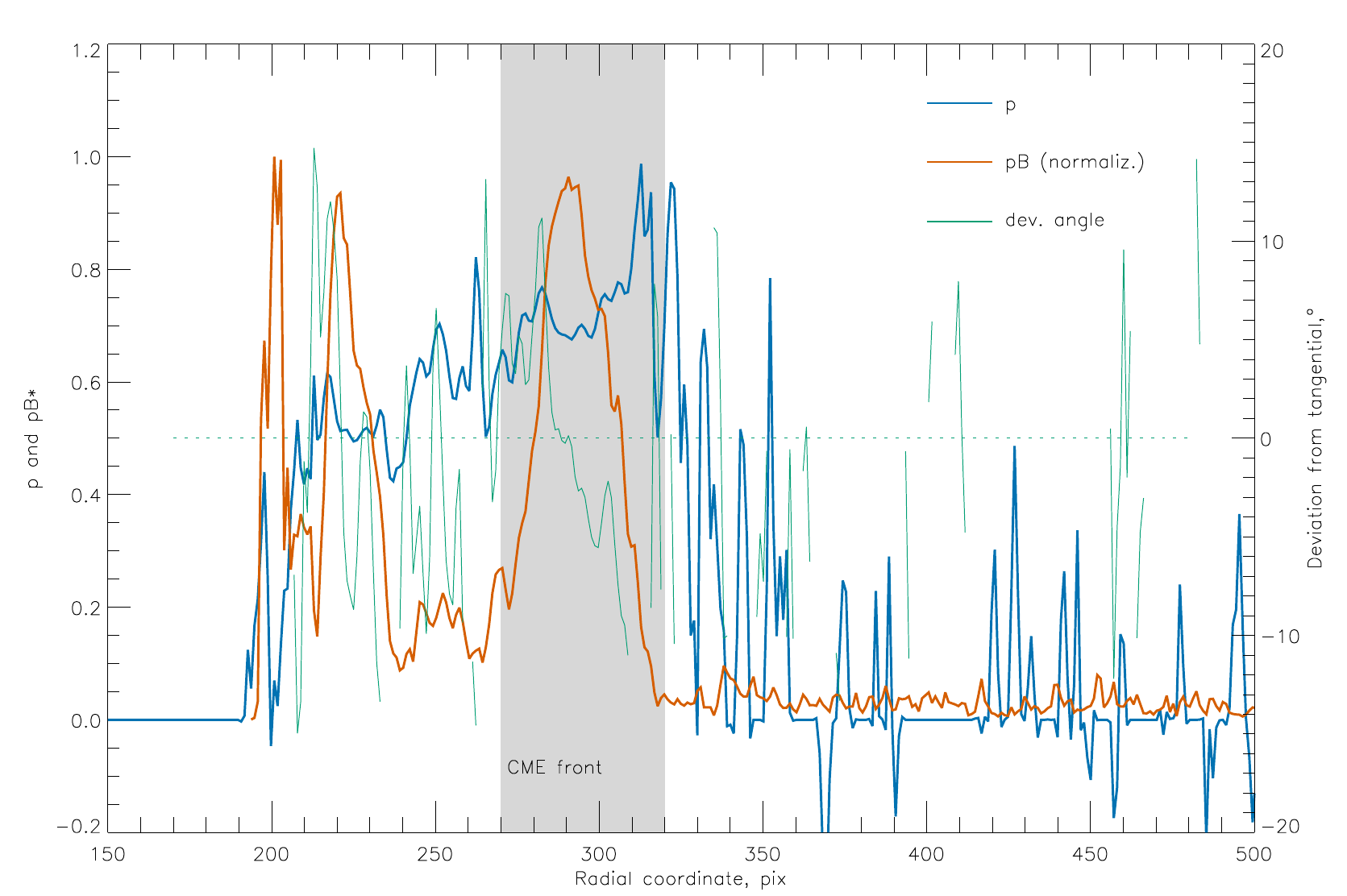}
    \caption{Radial profiles of the polarisation parameters along the dashed line in Figure~\ref{F-polarizparam}.} \label{F-profiles}
\end{figure}

We observe deviations from tangential greater than $10^\circ$, in regions associated with the excess brightness of the CME (see also Figure~\ref{F-profiles}). The degree of polarisation is around 70\,\% in regions associated with the CME signal (see also Figure~\ref{F-profiles}).

We apply the same technique to COR1-B and COR2-A and -B images, to see if the same behavior is observed in different instruments. 

\begin{figure}	
    \centering
   \includegraphics[width=0.9\textwidth]{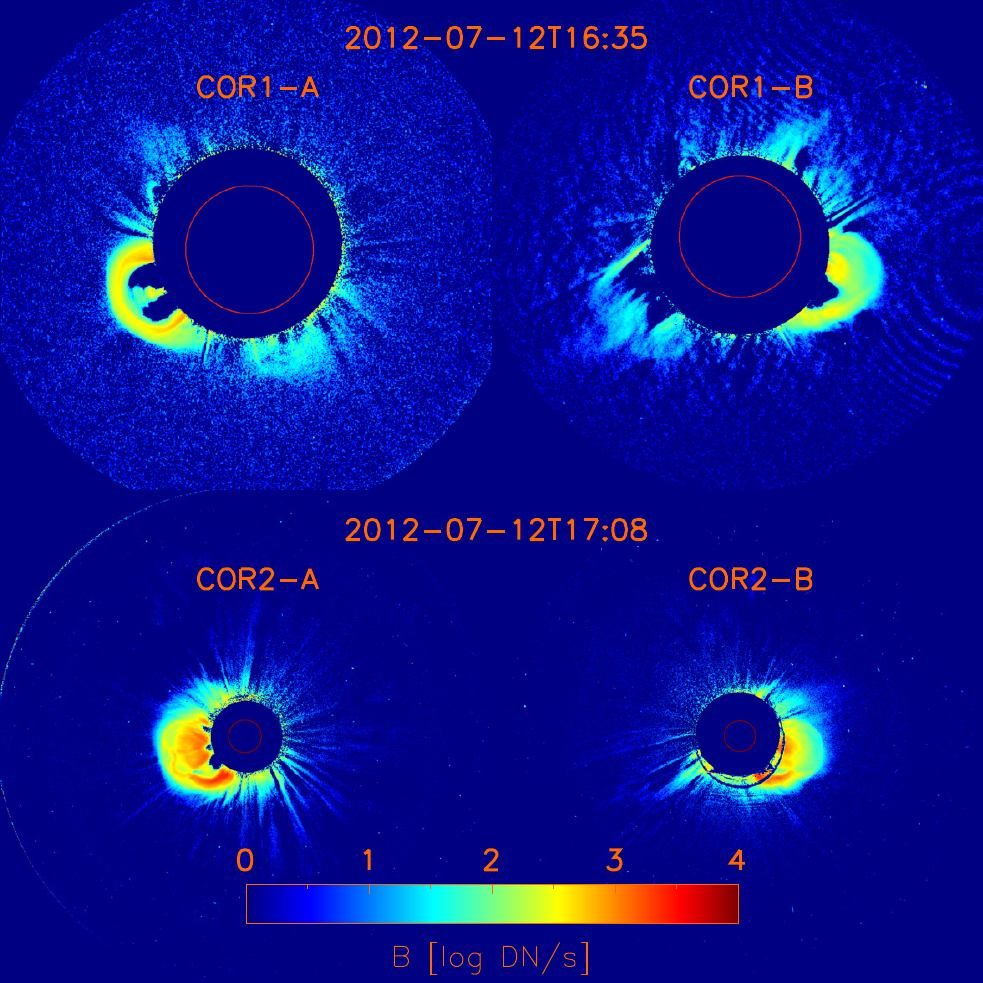}	
    \caption{Total brightness from COR1-A (\textit{upper-left panel}), COR1-B (\textit{upper-right panel}), COR2-A (\textit{lower-left panel}) and COR2-B (\textit{lower-right panel}) observations. The images are displayed in logarithmic scale. Base-difference images were used. The circle inside the occulter outlines the solar disk.} \label{F-totb}
\end{figure}

Figure~\ref{F-totb} displays the total brightness calculated from COR1-A (upper-left panel), COR1-B (upper-right panel), COR2-A (lower-left panel), and COR2-B (lower-right panel) images. The images are shown on a logarithmic scale. The two COR-B images show rather consistent structure of the CME, whereas the COR2-A structure is much more complex than the COR1-A one.

\begin{figure}	
    \centering
  \includegraphics[width=0.9\textwidth]{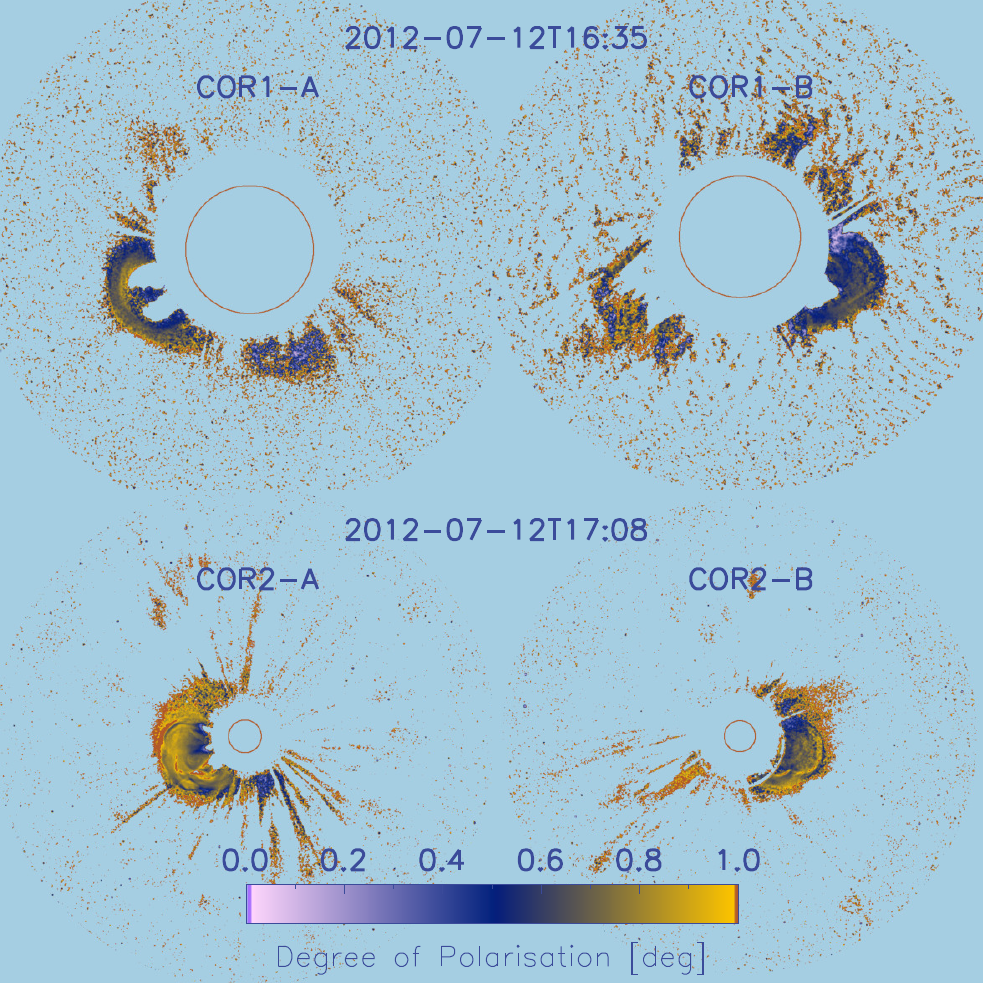}	
    \caption{Degree of polarisation from COR1-A (\textit{upper-left panel}), COR1-B (\textit{upper-right panel}), COR2-A (\textit{lower-left panel}), and COR2-B (\textit{lower-right panel}) observations. Base-difference images were used. The circle inside the occulter outlines the solar disk.} \label{F-degpol}
\end{figure}

Figure~\ref{F-degpol} shows the degree of polarisation from COR1-A (upper-left panel), COR1-B (upper-right panel), COR2-A (lower-left panel), and COR2-B (lower-right panel). 

The degree of polarisation is around 65\,\% but it can go as high as 80\,\% in some CME regions from COR1-A.
These high values of degree of polarisation are more observed in COR2 images.

\begin{figure}	
    \centering
   \includegraphics[width=0.9\textwidth]{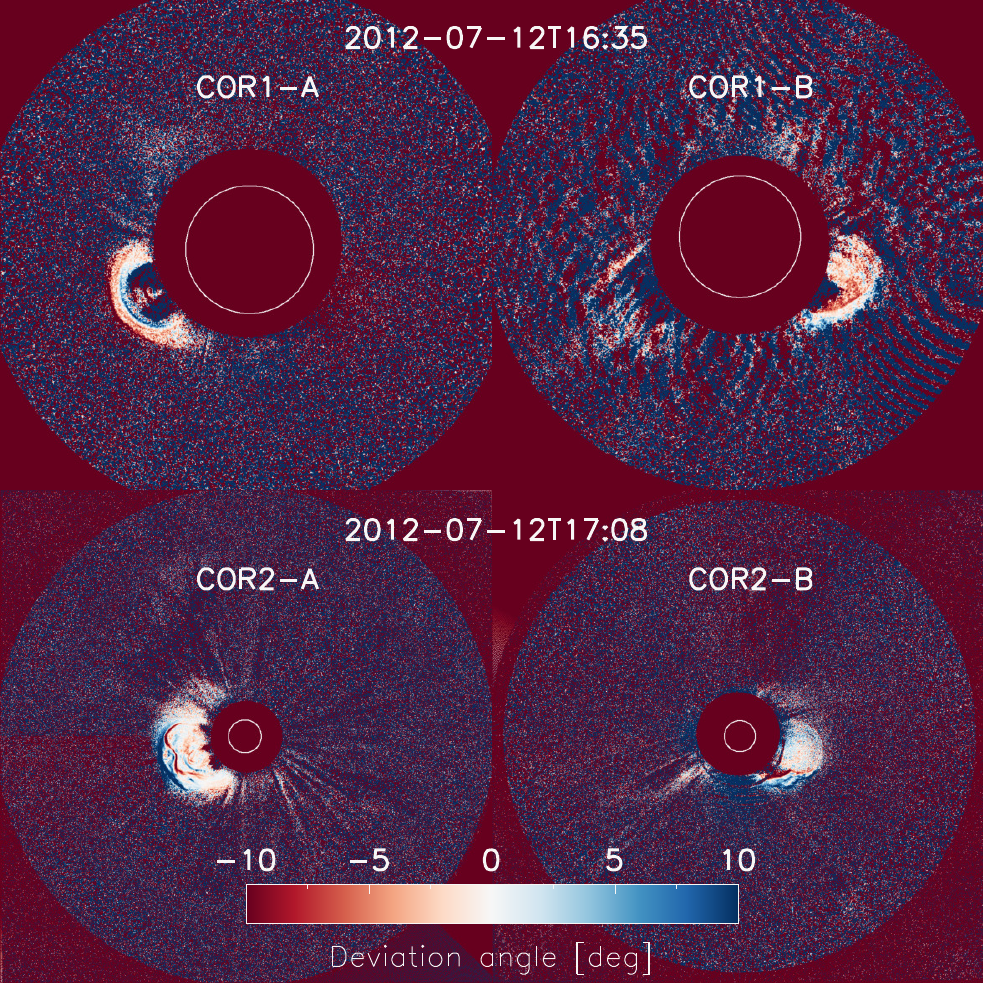}
    \caption{Deviation angle from tangential from COR1-A (\textit{upper-left panel}), COR1-B (\textit{upper-right panel}), COR2-A (\textit{lower-left panel}) and COR2-B (\textit{lower-right panel}) observations. Base-difference images were used. The circle inside the occulter outlines the solar disk.} \label{F-devang}
\end{figure}

In Figure~\ref{F-devang} is shown the deviation angle from tangential using COR1-A (upper-left panel), COR1-B (upper-right panel), COR2-A  (lower-left panel), and COR2-B (lower-right panel) data.

Deviations from tangential up to $10^\circ$--$15^\circ$ are seen in COR1-A images (upper-left panel of Figure~\ref{F-devang}), in all pixels where a significant CME excess brightness is observed, i.e. in both the front and the core of the CME (see also Figure~\ref{F-polarizparam}). It is more visible on the front part of the CME (the red and blue areas in the front and back part of the CME frontal structure). The deviations are smaller in COR1-B images. Same deviations are observed also in monthly-minimum and hourly-minimum background subtracted images, as discussed in Appendix~\ref{S-bkginfl}. For COR1, a systematic constant deviation of the polarisation angle by $-1^\circ$ was found \citep{Inhester2021}.

For COR2-A data, the deviation from tangential goes up to $30^\circ$ at the front of the CME (blue area in the lower-left panel of Figure~\ref{F-devang}).

\subsection{Tests on Deviation from Tangential} 
  \label{S-devang}
  
\begin{figure}	
    \centering
    \includegraphics[width=0.9\textwidth]{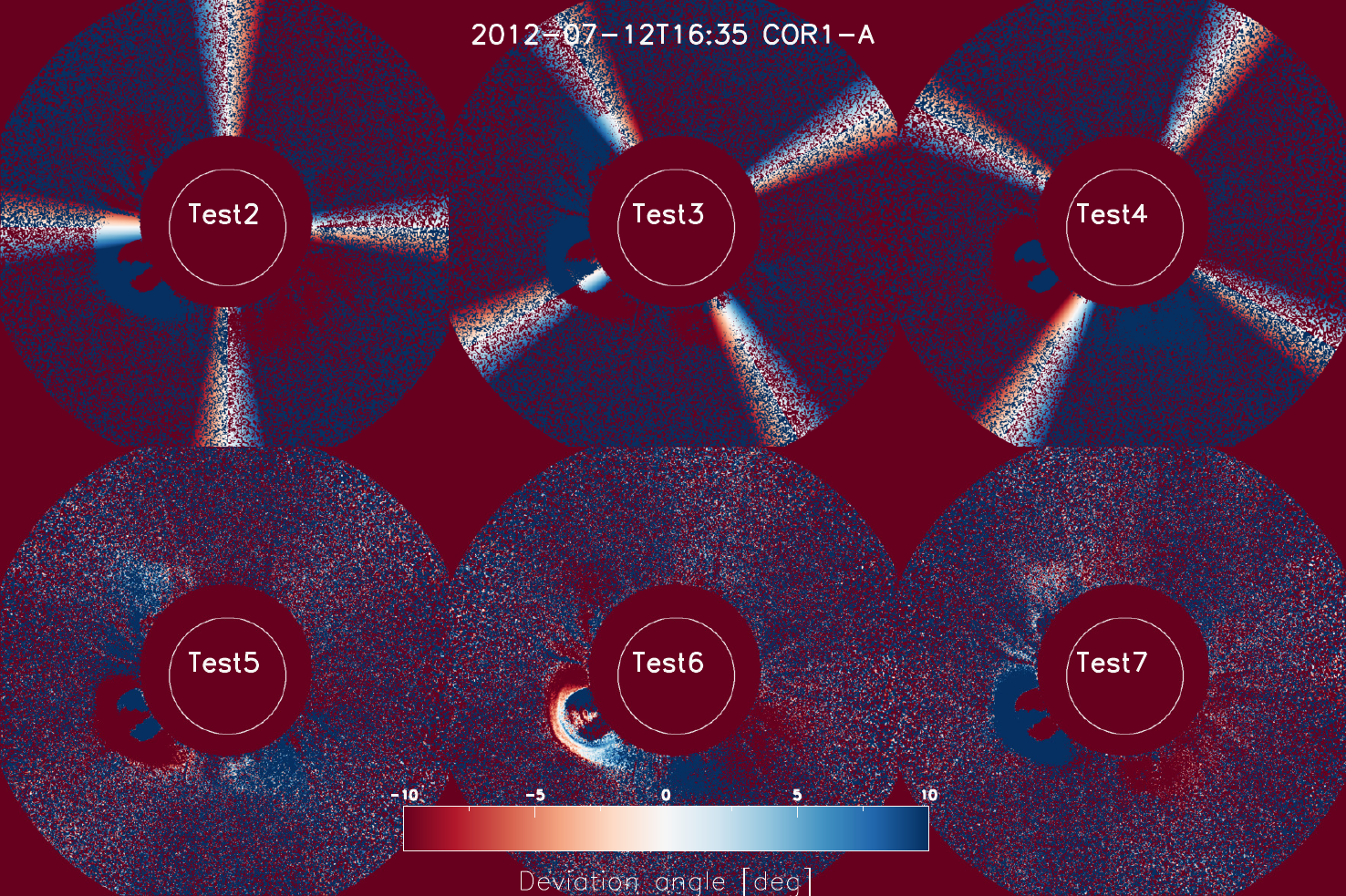}
    \caption{Deviation angle from tangential calculated in six different scenarios: \textit{Upper row}, from left to right: $I_{120}$ and $I_{240}$ set to zero, $I_{0}$ as observed; $I_{0}$ and $I_{240}$ set to zero, $I_{120}$ as observed; $I_{0}$ and $I_{120}$ set to zero and $I_{240}$ as observed. \textit{Lower row}, from left to right: $I_{0}$ set to zero and $I_{120}$ and $I_{240}$ as observed; $I_{120}$ set to zero and $I_{0}$ and $I_{240}$ as observed; $I_{240}$ set to zero and $I_{0}$ and $I_{120}$ as observed. The circle inside the occulter outlines the solar disk.} \label{F-devangtest}
\end{figure}

To further investigate the deviations from tangential described above, we perform a few tests as displayed in Figure~\ref{F-devangtest}. We used base-difference COR1-A images, at 16:35 UT for these tests.
\begin{enumerate}[i)]
\item In the first test (upper-left panel in Figure~\ref{F-devang}), we derive the polarisation angle from the three polarised images ($I_{0}$, $I_{120}$, $I_{240}$) recorded by COR1-A at 16:35 UT.
\item For the second test (left image of the upper row in Figure~\ref{F-devangtest}), we kept $I_{0}$ as observed and assigned $I_{120}$ and $I_{240}$ a zero value in all pixels.
\item For the third test (middle image of the upper row in Figure~\ref{F-devangtest}), we kept $I_{120}$ as observed and assigned $I_{0}$ and $I_{240}$ a zero value in all pixels.
\item For the fourth test (right image of the upper row in Figure~\ref{F-devangtest}), we kept $I_{240}$ as observed and assigned $I_{0}$ and $I_{120}$ a zero value in all pixels.
\item For the fifth test (left image of the lower row in Figure~\ref{F-devangtest}), we kept $I_{120}$ and $I_{240}$ as observed and assigned $I_{0}$ a zero value in all pixels.
\item For the sixth test (middle image of the lower row in Figure~\ref{F-devangtest}), we kept $I_{0}$ and $I_{240}$ as observed and assigned $I_{120}$ a zero value in all pixels.
\item For the seventh test (right image of the lower row in Figure~\ref{F-devangtest}), we kept $I_{0}$ and $I_{120}$ as observed and assigned $I_{240}$ a zero value in all pixels.
\end{enumerate}

We see that the observed deviation angles are almost entirely due only to the combination of $I_{0}$ and $I_{240}$ images, which are taken 18 seconds apart. In this time interval, a CME with a speed of 600\,km\,s$^{-1}$ will move about 10800\,km, i.e. more than two pixels (= 10489.8\,km) in COR1-A FOV, which may justify the deviations observed. The speed of 600\,km\,s$^{-1}$ was the measured speed of the CME at 2.41\,R$_{\odot}$ (16:35 UT) in the COR1-A FOV. 

In the next future, this interpretation will be further tested based on the analysis of synthetic CME observations built starting from a full MHD 3D simulation, similar to what was recently done by Bemporad, Pagano and Giordano (2018).

The values of the displacement of the CME during a polarisation sequence are summarised in Table~\ref{T-displacement}.

\begin{table}[h]
\caption{ The displacement of the CME during a polarisation sequence. First column shows the instrument taking the images, second column displays the time at which a sequence of three polarised images was taken, third column shows the speed of the CME in km s$^{-1}$, fourth column show the distance the CME traveled during the time interval the sequence was taken, which is displayed in column 5. Last column shows the displacement of the CME, in pixels, during the time interval that a sequence of three polarised images was taken.}
\label{T-displacement}
\begin{tabular}{cccccc}    
  \hline                   
Instrument & Time & Speed & Distance & Duration & Displacement \\
 & [hh:mm] & [km s$^{-1}$] & [km] & [s] & [pixels] \\
  \hline
COR1-A & 16:35  & 600 & 10800 & 18 & $>$2  \\
COR1-B & 16:35 & 700 & 16800 & 24 & $>$2 \\
COR2-A & 17:08 & 1200 & 72000 & 60 & $>$3 \\
COR2-B & 17:08 & 1250 & 75000 & 60 & $>$3 \\
  \hline
\end{tabular}
\end{table}

\subsection{CME Kinematics} 
  \label{S-kinem}
We calculated the projected speeds of the CME's leading edge observed by COR1 and COR2 at its central position angle (PA). The PA is measured from the North Pole, counterclockwise. For LASCO-C2, where the CME was observed as a full halo, we take the values from LASCO CME catalogue: \url{https://cdaw.gsfc.nasa.gov/CME\_list/}. The speed was measured at the PA corresponding to the fastest-moving segment of the CME leading edge. 

The values of the speeds as derived from different datasets are as follows:
\begin{itemize}
\item LASCO: speed of 885\,km s$^{-1}$ (linear fit), at a PA of 158$^{\circ}$. From a second order polynomial fit, the speed goes up to 1100\,km s$^{-1}$ at 5.5\,R$_{\odot}$.
\item COR1-A: speed from 200\,km s$^{-1}$ at around 2\,R$_{\odot}$ (16:20 UT) to 1100\,km s$^{-1}$ at around 3.9\,R$_{\odot}$ (16:55 UT), PA = 107$^{\circ}$.
\item COR1-B: speed from 500\,km s$^{-1}$ at around 2\,R$_{\odot}$ (16:25 UT) to 900\,km s$^{-1}$ at around 3.2\,R$_{\odot}$ (16:45 UT), PA = 254$^{\circ}$.
\item COR2-A: speed of 1000\,km s$^{-1}$, at PA = 107$^{\circ}$, and height of 14\,R$_{\odot}$  (18:39 UT) (decelerating from 1200\,km s$^{-1}$).
\item COR2-B: speed of 1250\,km s$^{-1}$, at PA = 254$^{\circ}$, and height of 14\,R$_{\odot}$ (18:24 UT) (slightly decelerating).

\end{itemize}

\begin{figure}	
    \centering
   \includegraphics[width=0.9\textwidth]{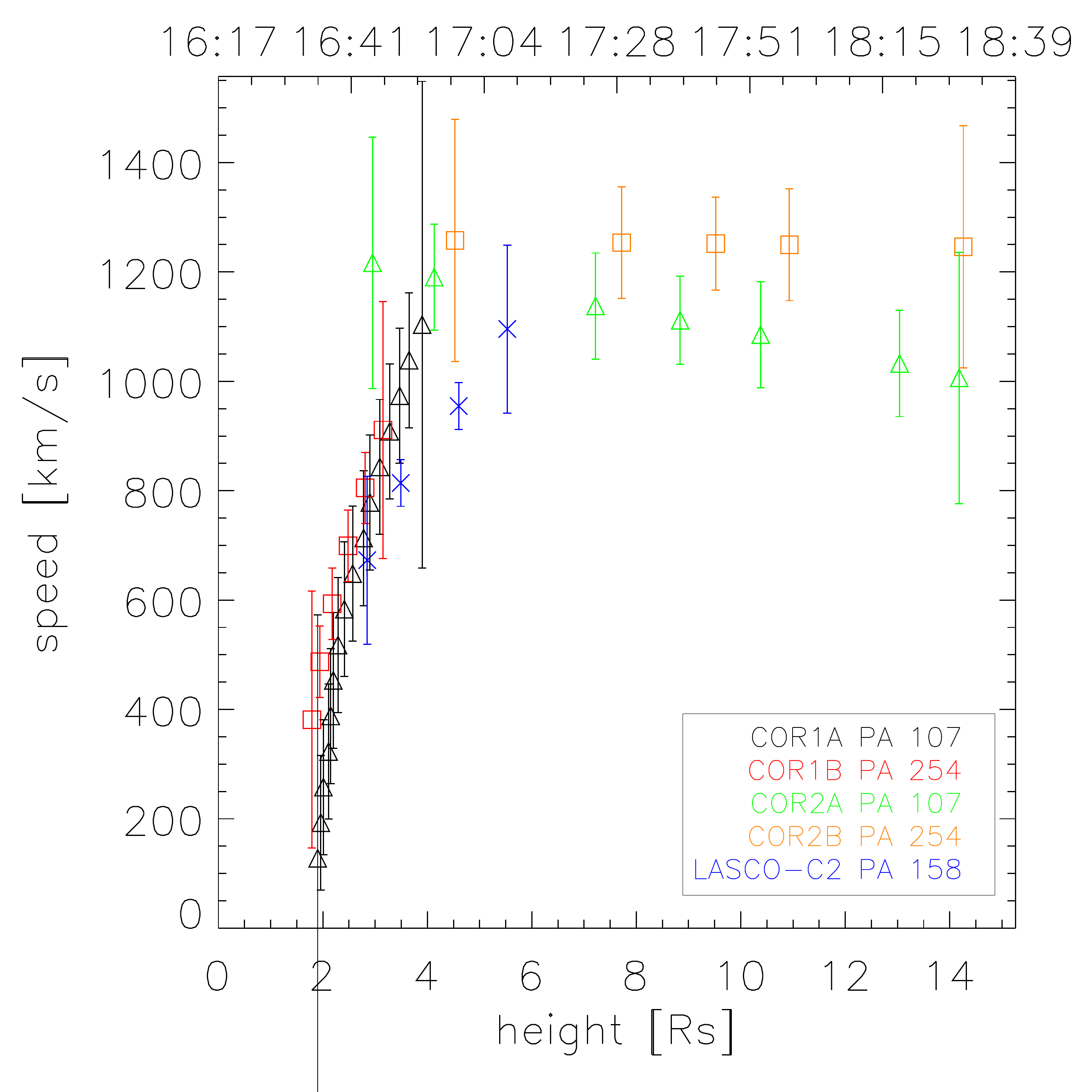}	
    \caption{Projected speeds versus height as measured in COR1-A at PA $107^\circ$ (\textit{black triangles}), COR1-B at PA 254$^\circ$ (\textit{red squares}), COR2-A at 107$^\circ$ (\textit{green triangles}), COR2-B at PA 254$^\circ$ (\textit{orange squares}), and LASCO-C2 at 158$^\circ$ (\textit{blue X-s}). The errors are derived by assuming an error in height of five pixels.} \label{F-speeds}
\end{figure}

These values are shown also in Figure~\ref{F-speeds}. 

It is observed that the CME accelerates quite a lot in the regions of the corona below 4 R$_{\odot}$ and keeps approximately a constant speed above 4 R$_{\odot}$.

For a COR1 pixel size of 7.5$^{''}$, i.e. 5245\,km\,pixel$^{-1}$ (COR1-A) and 5527\,km\,pixel$^{-1}$ (COR1-B), a CME moving at 1000\,km s$^{-1}$ will cross the single pixel in 5.24 seconds (COR1-A) and 5.53 seconds (COR1-B). This is larger than the nominal exposure time of one second. For a COR2 pixel size of 29.4$^{''}$, i.e. 20549\,km\,pixel$^{-1}$ (COR2-A) and 21662\,km\,pixel$^{-1}$ (COR2-B), a CME moving at 1000\,km s$^{-1}$ will cross the single pixel in 20.55 seconds (COR2-A) and 21.66 seconds (COR2-B), which is again larger than the nominal exposure time of six seconds. Hence, plasma displacements during the exposure times can be neglected.

\subsection{CME 3D Reconstruction} 
  \label{S-3dreconstr}

We applied three different reconstruction techniques in order to derive the 3D position of the CME, namely: triangulation \citep{Inhester2006}, polarisation ratio (PR) \citep[see e.g.,][]{Moran2004, Dere2005, Vourlidas2006, Floyd2019}, and graduated cylindrical shell (GCS) model \citep[see][]{Thernisien2011}.

\begin{figure}	
    \centering
  \includegraphics[width=0.3\textwidth]{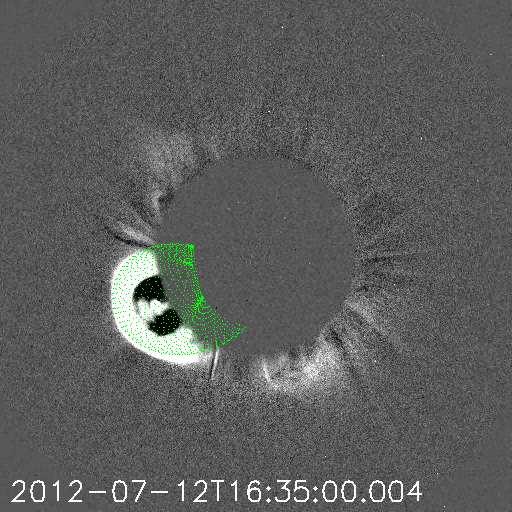}
  \includegraphics[width=0.3\textwidth]{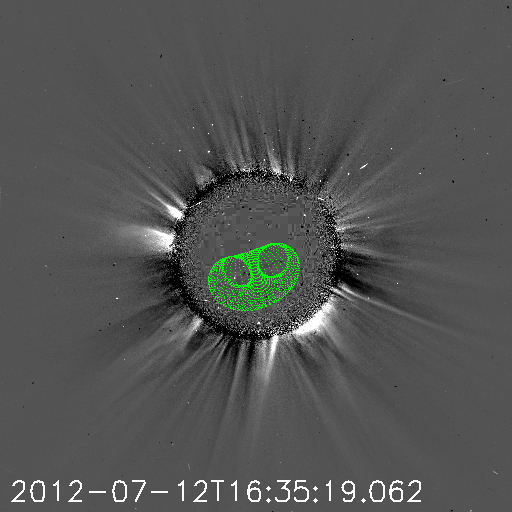}
  \includegraphics[width=0.3\textwidth]{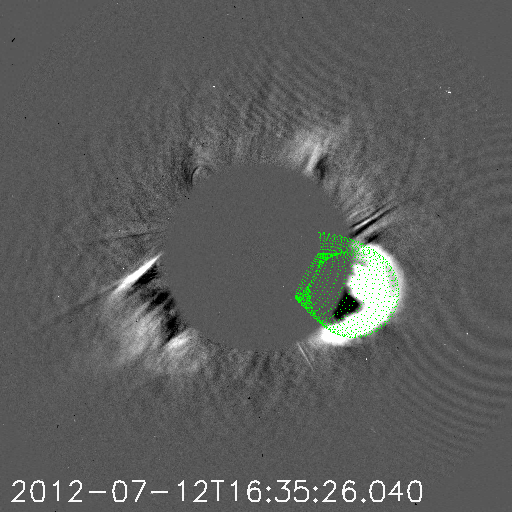}
  \includegraphics[width=0.3\textwidth]{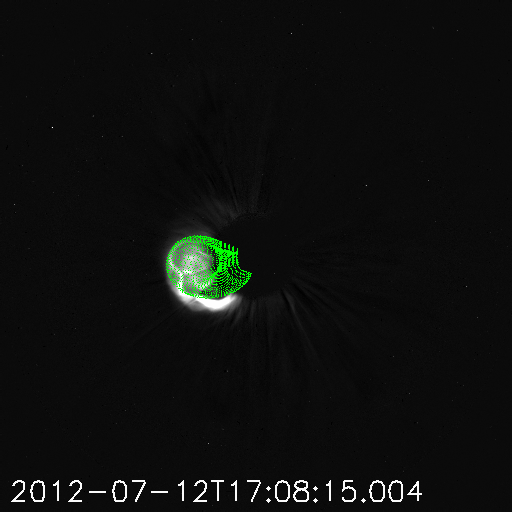}
  \includegraphics[width=0.3\textwidth]{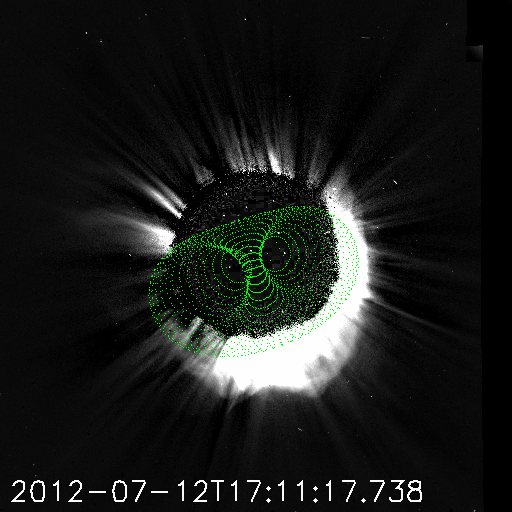}
  \includegraphics[width=0.3\textwidth]{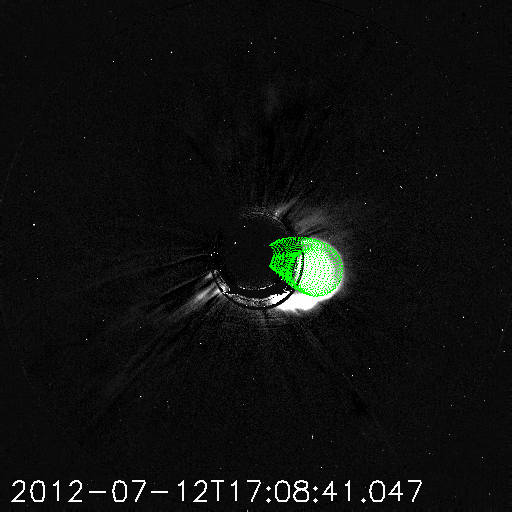}
    \caption{GCS reconstruction for the CME observed by STEREO/COR-A (\textit{left panels}), LASCO-C2 (\textit{middle panels}) and STEREO/COR-B (\textit{right panels}) at around 16:35 UT (\textit{upper panels}), and 17:08 (\textit{lower panels}) on 12 July 2012.} \label{F-gcs}
\end{figure}

The polarisation-ratio technique was applied on a small region at the central part of the CME leading edge and on the core of the CME as seen in COR1-A at 16:35 UT. Triangulation was applied on the same regions.
GCS was applied on the CME observed by COR1 at 16:35 and by COR2 at 17:08 UT. The results of GCS are displayed in Figure~\ref{F-gcs}.

\begin{figure}	
    \centering
  \includegraphics[width=0.9\textwidth]{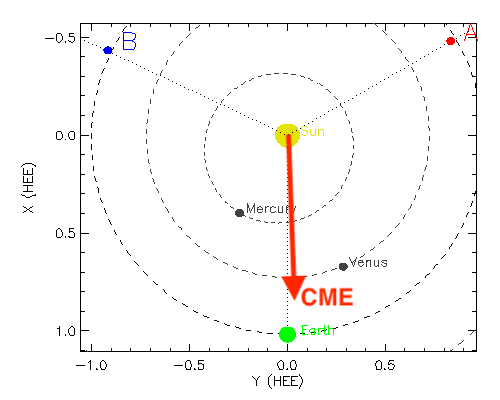}	
    \caption{Position of the three spacecraft (STEREO-A (\textit{red}), STEREO-B (\textit{blue}), and SOHO - Earth perspective (\textit{green}) on 12 July 2012, 16:00). The red arrow indicates the direction of propagation of the CME as derived from GCS reconstruction. The view is from above, in the Heliocentric Earth Ecliptic (HEE) coordinate system, with $x$-axis towards the Earth and its $z$-axis towards Ecliptic North Pole \citep{Thompson2006}.} \label{F-position}
\end{figure}

The results from different reconstruction techniques are summarised in Table~\ref{T-cmeposition} together with the location of STEREO-A, STEREO-B, and Earth on 12 July 2012, 16:00 (also shown in Figure~\ref{F-position}). 

It is observed that the three methods, GCS, triangulation, and polarised ratio method, yield the same values for the position of the scattering material at distances around 2\,R$_{\odot}$. At higher distances (around 6\,R$_{\odot}$) the CME is slightly deflecting towards the Equator.

Given the position of the spacecraft and the direction of propagation of the CME, this was observed at i) the central meridian as seen from the Earth; ii) $30^\circ$ behind the East limb as observed by STEREO-A and iii) $25^\circ$ behind the West limb as observed by STEREO-B.

This is also visible in EUVI\,195 -A and -B data where dimmings appear as being located behind the limb (see, e.g., \url{https://cdaw.gsfc.nasa.gov/stereo/daily\_movies/2012/07/12/index.html}).

\begin{table}[!h]
\caption{3D position of the CME as derived from different reconstruction techniques \citep[in HEEQ coordinates; see][]{Thompson2006}. The last three entries show the positions of the spacecraft, also in HEEQ. }
\label{T-cmeposition}
\begin{tabular}{cccccc}    
  \hline                   
Time & lat/lon & Height & Feature & Technique & Instruments \\
\hline
16:00 & S18W01 & - & on-disk source & - & AIA  \\
16:35 & S18W05 & 2.48 & LE & Triang & COR1  \\
16:35 & S17E03 & 2.06 & core & Triang & COR1  \\
16:35 & S18W05 & 2.75 & LE-A & PR & COR1  \\
16:35 & S16W05 & 2.15 & core-A & PR & COR1  \\
16:35 & S18W00 & 2.5 & LE & GCS & COR1 + C2  \\
17:08 & S07W00 & 5.7 & LE & GCS & COR2 + C2  \\
  \hline
16:00 & N04W00 & - & - & - & Earth \\
16:00 & N03W120 & - & - & - & STEREO-A \\
16:00 & S07E115 & - & - & - & STEREO-B \\
\hline
\end{tabular}
\end{table}

\section{Discussion} 
\label{S-Discussions} 

\subsection{Discussion on the Degree of Polarisation} 
\label{S-discussiondegpol}
The degree of polarisation of the K-corona, depends on the distance from the solar surface and on the scattering angle to the observer. Besides the position in space of the structure (streamer, coronal mass ejection, etc.), the effective line-of-sight integration length is also important when deducing the degree of polarisation \citep{Koutchmy1972}.

We have found that the degree of polarisation is around 65\,\% but it can go as high as 80\,\% in some CME regions from COR1-A.
\citet{Floyd2019} showed that for an electron in the POS at 2\,R$_{\odot}$, the polarisation is about 0.8 and increases to nearly 1 at 3\,R$_{\odot}$. These values decrease with increasing angle from the POS.

Similar or lower values of the degree of polarisation were observed for coronal streamers. \citet{Lamy2020} by studying polarised LASCO-C2 images showed that the brightest streamers culminate at a peak degree of polarisation of around 0.4 at 2.2\,R$_{\odot}$, at maximum of solar activity. \citet{Durst1982}, by studying the solar eclipse of 16 February 1980, found that the degree of polarisation of the K-corona is strongly increasing with height in the inner corona and it is almost constant for heights higher than 2\,R$_{\odot}$. Several streamers located in or near the POS have maximum polarisation near 80\,\%. \citet{Koutchmy1977}, by studying the eclipse of 30 June 1973, found polarisations as high as 60\,\% in regions associated with coronal streamers.
In another solar-eclipse study, \citet{Vorobiev2020} found that the degree of polarisation of the F+K corona (including also the sky) peaks at around 47\,\% in the equatorial features, at a distance of around 1.5\,R$_{\odot}$. 

An unexpectedly large degree of polarisation (exceeding the maximum non-relativistic Thomson scatter value) was observed by a number of authors \citep{Koutchmy1971, Pepin1970, Skomorovsky2012, Qu2013}. This was attributed by  \citet{Qu2013} to the contamination of the scattered white-light signal by coronal ion emission lines.
As the majority of these studies were made during solar eclipses one extra source of contamination may come from the sky polarisation \citep[see][]{Koutchmy1999}. Obviously this is not the case for space data. 

\subsection{Discussion on the Polarisation Angle} 
\label{S-discussionpolangle}
The anisotropy of the incident light causes the observed scattered radiation to exhibit a polarisation parallel to the visible solar limb (tangential to the solar limb).

In our case we observe deviations from tangential higher than $10^\circ$, in regions associated with the CME brightness excess.
\citet{Inhester2021} analysed COR1 observations of the quiescent corona and found a systematic constant deviation of the polarisation angle by $-1^\circ$, which exceeded the calculated variance of the polarisation angle. As the most simple and most probable explanation the authors proposed not-precise-enough knowledge of orientation of the polarisers and thus an incorrect demodulation matrix. The offset of $2.5^\circ$ in mechanical mounting of the polarisers in COR1 would provide systematic deviation of order of $1^\circ$. Such an explanation is not applicable in our case, since the polarisation angle varies across the CME and takes larger values. We look for other possible explanation in the following sections.

\subsection{Possible Interpretation of the Observed Deviation Angles: Relativistic Electrons?} 
\label{S-relativistic}

It have been shown that deviation from tangential may occur and among possible physical reasons of such deviations we mention the Thomson scattering by the fast moving (relativistic) electrons in the solar corona and the scattering by the supra-thermal electrons \citep{Molodensky1973, Kishonkov1975}. These kinds of deviations have been reported only seldom in the literature \citep{Pepin1970, Badalyan1988, Skomorovsky2012, Qu2013} and some results deduced from photographic films/plates are of limited precision and significance \citep{Park2001, Kim2015}, especially when radial neutral filters were not used. Other studies explicitly found no deviation from tangential polarisation within the measurement error \citep{Koutchmy1993, Filippov1994, Kim1996, Kulijanishvili2005, Vorobiev2020}.

In the case of a CME-driven shock wave the semi-relativistic electrons can be expected at the front or the flanks of the CME inducing radio emission (type II and type III radio bursts) \citep[see, e.g.,][]{Classen2002, Magdalenic2014}. The energies of impulsive electron events near the Sun could be in the range of 1-40 keV (sometimes going down to 0.1 keV) \citep{Reid2014}. 
A small, e.g. order of 10\,\% population of high energy electrons is sufficient to produce bursts. For a net rotation of the polarisation angle the contribution of all electrons adds up incoherently. So if there are 10\,\% beamed high energy electrons their scatter with rotated polarisation adds to the unrotated Thomson background of the other 90\,\%. This may strongly reduce the observed net rotation.
If the electrons are not beamed but just isotropically heated, then each energetic electron in  the tail of the distribution may rotate its scatter but if all contributions are added up this results not in a net observable rotation of the polarisation direction but just in a random broadening of the polarisation directions which probably is observed as a slightly reduced polarisation degree.

Small relativistic effects may be observed even for an electron energy of 0.03 keV with deviations relative to the non-relativistic limit of a few \% for the degree of polarisation and 2 degrees away from the tangential direction for the polarisation angle \citep{Inhester2015}.

\subsection{Possible Interpretation of the Observed Deviation Angles: Line Contamination?} 
  \label{S-linecontamin}
  
The visible emission lines (H$\alpha$, Helium D3-line, Fe\,{\sc xiv}, Fe\,{\sc x}, etc.) are part of the signal of a coronagraph image together with the K-corona, F-corona and the stray-light. 

As described in Section~\ref{S-data}, COR1 observes in a white-light waveband 22.5\,nm wide centred at the H$\alpha$ line at 656\,nm, and COR2 observes in visible light with a passband from 650 to 750\,nm. This implies that the CME signal observed by COR1 may be ''contaminated'' with H$\alpha$ emission.

To investigate potential contribution from H$\alpha$ emission, we applied the polarisation ratio technique on COR1-A data at 16:35. If the H$\alpha$ contribution is significant we should observe low polarisation in the region of our CME \citep[see, e.g.,][]{Poland1976, Mierla2011}. Instead, from Figure~\ref{F-profiles} it is clear that a high degree of polarisation of order of 0.7 is observed in the CME region. 
Note also that the H$\alpha$ emission is supposed to be observed in the core of the CME, but the deviation from the tangential was most visible in the leading edge.

Let us compare our reasoning with other events. As was pointed out by \citet{Floyd2019}, the Helium D3 line at 587.7\,nm, falls right at the center of the LASCO-C2 orange bandpass and it was seen in the core of many three-part CMEs imaged by C2. The authors concluded that the prominence material embedded in LASCO CMEs is expected to have a maximum polarisation of the order of 17\,\% and a direction somewhat deviating from tangential.
The same conclusion was derived from another study by \citet{Heinzel2020}. In quiescent prominences, the polarisation of the He\,{\sc I} D3 emission is low, typically 2\,\% and its direction deviates from tangential by approximately $\pm15^\circ$ (Leroy, Ratier, and Bommier, 1977).

\cite{Dolei2014}, by studying three blobs observed in COR1 images, demonstrated that the H$\alpha$ contribution to the total white-light intensity observed at the location of the blobs is between 95 and 98\,\%. They also showed that the H$\alpha$ polarised component is of the order of a few percent of the H$\alpha$ total emission from the blobs.

\citet{Mierla2011} demonstrated that the major part of a CME core emission (more than 85\,\%), is H$\alpha$ radiation, and only a small fraction is Thomson-scattered light. This results in the reduction of the CME polarisation signal. The CME was associated with a filament eruption, and it was observed by the COR1 instrument.

The same was deduced from the eclipse of 2019 with the CATE experiment (D. Elmore, private communication). They observed a prominence at NNE limb with polarisation close to zero over the polarised coronal background.

For this study EUV images in 304 and 195\,$\AA$ taken by EUVI and AIA were inspected and no filament/prominence was found. Nevertheless, the CME had a three-part structure with a bright leading edge, followed by a dark cavity and a bright core, meaning it was a flux-rope-like ejecta. Such CMEs without any associated filaments are often reported in the literature, e.g. \citet{Howard2017, Vourlidas2013}. From the 3D reconstruction described above we saw that the two methods, triangulation and polarisation ratio method, yield the same values for the position of the scattering material.

Thus we can conclude that there is no H$\alpha$ emission associated with our CME, i.e. the signal of the CME is due entirely to the Thomson-scattered light.

\subsection{Other Effects That Can Modify the Polarisation Angle and the Degree of Polarisation} 
  \label{S-othereff}

Other phenomena that can create low polarisation in coronagraph images are F-corona emission \citep[e.g.,][]{Morgan2007}, and Thomson scattering from enhanced plasma density far away from the POS \citep[e.g.,][]{Billings1966}. F-corona contribution is ruled out because it forms a diffuse background and does not vary rapidly in time. The subtraction of the image taken 45 minutes before our image at 16:35 should have removed any F-corona contribution. 

As shown above, the CME appears at different locations in each polarised image resulting in an error in the measured polarisation. This effect may be taken care of by binning the images. \citet{Thernisien2011b} showed that substantial binning of the LASCO-C2 images would be required to suppress the effect, for instance by a factor of about 36 (from the original format of $1024\times1024$ pixels) for a CME traveling at 500\,km s$^{-1}$. 

We performed a similar test with the COR1-A image at 16:35 UT. By binning the image to a size of $64\times64$ (i.e. 16 times less than the original image) we noticed that the deviation angle from tangential is around $0\pm3^\circ$, while in the binned $128\times128$ images one can still see small deviations (around $7^\circ$), especially in the front part of the CME. In these regions, the original image can have deviation as high as $15^\circ$ or more (see Figure~\ref{F-profiles}). The binned images of $32\times32$ were too noisy to conclude anything.

The degree of polarisation for the binned images behaves in a similar way to the deviation angle images, i.e. the high values (above 0.65) decrease or disappear with the size of the image. 

\subsection{Discussion on Different Effects Influencing the Polarisation Analysis} 
  \label{S-effectsdiscussion}
  
In a recent study by \citet{Floyd2019}, the main limitations the authors encountered when deriving the polarisation parameters of 15 CMEs observed by LASCO-C2 were: i) the contamination of the LASCO-C2 passband filter with the He\,{\sc I} D3 line at 587.7\,nm. 
ii) the duration of ten minutes of a LASCO-C2 polarisation sequence resulting in the blurring of the polarisation in case of the fast CMEs. The front of fast CMEs appeared at different locations in each polarised image resulting in an error in the measured polarisation. 
iii) using a pre-event image resulted in artifacts affecting the \textit{pB}-images of the LASCO-C2 CMEs because traces of underlying coronal structures are interfering with the CME signal.

In another study, \citet{Lamy2020}, using the newest calibrated LASCO-C2 polarised images, analysed the polarised K-corona over the two complete Solar Cycles 23 and 24, and they compare it with a variety of eclipse data taken during the two solar cycles. The agreement with the eclipse data is in general good except for slight discrepancies affecting the innermost part of the C2 field of view.

To do this analysis, the authors needed to introduce further corrections to the polarised data: 
i) the correction for the global transmission of the $0^\circ$ polariser; 
ii) global correction function applied to polarised images; 
iii) exposure-time equalization of the three polarised images. 
They showed that errors on the relative global transmission between polarisers and on the individual exposure times of the triplet forming a polarisation sequence have similar effects, which were corrected by an optimization procedure based on imposing a tangential direction to the solar limb of polarisation and minimal dispersion around this direction. 
They concluded that capability of LASCO-C2 to perform accurate polarimetric measurements is affected by several adverse conditions, prominently the very basic polarisation analyzer system (three Polaroid foils), the optical design with two folding mirrors, and the fluctuations in exposure times.

A similar study was performed by \citet{Lamy2021}, this time with the newest calibrated LASCO-C3 polarised data. They only used two polarisers in this case, because of the degradation of the $0^{\circ}$ polariser.

In the case of internally occulted coronagraphs (like COR1) multiple reflections in the instrument from the direct photospheric light are difficult to avoid. When using an externally occulted instrument simulating a total eclipse (LASCO-C2, COR2, etc.), the photospheric light does not reach the instrument, but several new problems appear as described by \citet{Lamy2020}. Some significant improvement was made in the case of the COR1 and COR2 coronagraphs by using a rotating single polariser instead of comparing images taken using different polarisers, as done with LASCO-C2. However, in the case of COR1 and COR2 although polarisation images are taken through the same optics, the rotation of the polariser introduces some wobble between the images due to the fact that the front and rear surfaces of the polariser are not exactly parallel. This effect is smaller than a pixel, but the slight misregistration between the three images can produce some spurious polarisation signals at sharp transitions in brightness (see e.g. the analysis of the Comet Lovejoy by \citet{Thompson2015}). This is definitely an issue when analyzing comet data, but is not generally a problem for the solar corona, where the variations in brightness tend to be more gradual. In our case the effect is negligible compared to other sources of uncertainty.

To summarise, the causes affecting the polarimetric measurements may be related to 
i) the instrument design: e.g. the optical design with two folding mirrors as in the case of LASCO-C2; the basic polarisation analyzer system (three Polaroid foils as in LASCO-C2, and one polariser which rotates as in COR).
ii) the instrumental operation mode: e.g. the fluctuations in exposure times; the time difference in taking a polarisation sequence (ten minutes for LASCO-C2, a few seconds for COR1 and COR2). 
iii) the correct extraction of the K-corona signal from other signals (different background subtraction are involved).
iv) contamination of the bandpass with emission lines (He I D3 for LASCO-C2 and H$\alpha$ for COR1). 

From the effects specified above, the one affecting the most our measurements is related to the time span in taking a polarisation sequence. The other effects are minimised by our data pre-processing (with the \textsf{secchi\_prep} routine), by background subtraction (to keep only the CME K-corona signal - see discussions in Appendix~\ref{S-bkginfl}) and by the type of the observation (CME without any prominence associated, i.e. no emission-line contamination).

\subsection{Improving the Polarisation Analysis} 
  \label{S-improvements}
  
The questions arises of how we can improve polarisation measurements of CMEs. Of course one aspect is low scattered light, but this is not the key aspect since scattered light is subtracted by the pre-event image. Key areas of improvement are the accuracy, hence the technique (three polaroids being the worst) and the speed of the measurements (to ''freeze'' the motion of the CMEs).  

The \textit{Solar Orbiter visible light and ultraviolet coronal imager} (Metis) \citep{Antonucci2020} has the technical capability of obtaining polarisation measurements of quality superior to those of LASCO-C2 and COR1 thanks to its electro-optical device, which does not require any mechanical motion. The resulting much faster mode of operation will contribute to the reduction of the time delays between the individual images required to obtain the deviation-angle image.
  
One step forward to improve the derivation of the polarisation parameters was proposed by Vorobiev et al. 2020. The authors used a new polarimetric imaging camera and method: a set of four linear micro-polarisers at 0, 45, 90, and $135^\circ$ put over each 
''sub-matrix'' made of ''super pixels'' of 2$\times$2 assembled on the CCD chip. They could detect polarisation signals as small as 0.3\,\% and found no strong deviations in the angle of linear polarisation from the tangential direction.

\citet{Koutchmy2019} also performed such analyses by using a rotating linear Polaroid placed before the entrance lens of F300 mm at F/5.6 and a Canon D60 color CMOS camera of 24 Mpx. Twelve images were recorded at positions consecutively shifted by 15$^\circ$, with a precision better than one degree and a cadence of four seconds using the same exposure time 1/8 second. No deviation from the tangential direction was found (at the 1$^\circ$ level of precision) after subtracting the sky background as measured over the image of the Moon. 

Among the next space missions with new polarisation technical developments is BITSE - \textit{Balloon-borne Investigation of Temperature and Speed of Electrons} in the corona \citep[see][]{Gong2019}. The BITSE’s detector will be able to directly collect polarised light at four different wavelengths: 385.0, 398.7, 410.0, and 423.3\,nm. It will measure the coronal average electron temperature using the Grotrian method.

PUNCH - \textit{Polarimeter to UNify the Corona and Heliosphere} will also have polarisation capability. It comprises four matched cameras all operating as a ''virtual instrument'' to image Thomson-scattered light, from the vantage of four separate spacecraft in Sun-synchronous LEO. PUNCH is the first coronal and solar-wind imager designed specifically to produce three dimensional images from a single vantage point \citep{DeForest2019}.

The future externally occulted coronagraph ASPIICS (\textit{the Association of Spacecraft for Polarimetric and Imaging Investigation of the Corona of the Sun}), onboard PROBA-3 (\textit{the Project for On-Board Autonomy}), will be the first space-borne instrument to image the polarised corona at distances as low as 1.1\,R$_{\odot}$, with low straylight \citep{Lamy2010, Renotte2015, Galano2018}. The coronagraph will be equipped with a filter wheel with six positions to choose between three spectral filters and three linear polarisers. ASPIICS will not suffer from radial vignetting due to its remote external occulter and it will preserve its image quality throughout its FOV. Nevertheless, due to the usage of a filter wheel there will be limitations because of the non-simultaneity in taking the set of three polarised images.

\section{Conclusions} 
      \label{S-Conclusions} 

In this work we have used  SECCHI/COR1 and COR2 polarised white-light data to study a large and fast CME on 12 July 2012. We used the degree of polarisation to calculate the position of the CME in 3D. We compared it with the position derived from triangulation and forward modelling techniques and we found very good agreement among them.
From the study of the CME kinematics we derived projected speeds up to around 1000~km~s$^{-1}$.
The CME studied here showed deviation of the polarisation angle from the tangential in the range of $10\,--\,15^\circ$ (or more).
Our analysis showed that this is mostly due to the fact that the sequence of three polarised images from which the polarised parameters are derived is not taken simultaneously, but at a difference of few seconds in time. In this interval of time, the CME is moving by at least two pixels in the FOV of the instruments, and this displacement results in uncertainties in the polarisation parameters (degree of polarisation, polarisation angle, etc.).

We conclude that the observational effect of the propagation of the CME front through the image combined with non-simultaneous polarisation observations is the most likely reason of the deviation of the polarisation angle from the tangential direction. To our knowledge this effect has not been considered before.

This study is important also for analysing future data from instruments with polarisation capabilities.

\begin{acks}
 M. Mierla thanks Robin Colaninno and Angelos Vourlidas for fruitful discussions on SECCHI-COR backgrounds. She also thanks Jan Janssens for pointing out possible filament-related sources and to Bill Thompson for clarifying the issue with COR2 polarisation angle. The authors acknowledge the use of SECCHI-COR1, -COR2, and LASCO-C2 data. M. Mierla, A.N. Zhukov, and S.V. Shestov thank the Belgian Federal Science Policy Office (BELSPO) for the provision of financial support in the framework of the PRODEX Programme of the European Space Agency (ESA) under contract numbers 4000117262, 4000134474 and 4000136424.
\end{acks}

\begin{dataavailability}
All the data used in this study (SECCHI-COR1, -COR2, and LASCO-C2) are publicly available and can be downloaded from the corresponding instrument webpages.
\end{dataavailability}
 
\begin{ethics}
\begin{conflict}
The authors declare that they have no conflicts of interest.
\end{conflict}
\end{ethics}


\appendix

\section{Background Subtraction} \label{S-bkg}

In order to keep signal only from the K-corona one can employ different procedures on raw white-light coronagraph images: 

i) subtraction of the monthly minimum background from a raw white-light coronagraph image. The monthly minimum background is created by taking the minimum in each pixel over a four week period of daily median images. It removes the static K-corona, the F-corona, and the stray light from the images, assuming that the F-corona and the stray light do not change much during one month period. Overall, this procedure works well but it can remove also parts of the stable K-corona (and parts of F-corona and stray light may still be present in the image). \\

ii) subtraction of a previous image from a raw white-light coronagraph image. This is mostly used for studying K-corona dynamic features like CMEs. One can subtract a pre-event image (the so-called base difference image technique) or the precedent image (the so-called running-difference image technique) in order to keep only the signal from the dynamic feature we are interested in. Nevertheless, static features and the E-corona signal may still be present in the final image. The main assumption here is that the stray-light and F-corona does not vary over few minutes/hours.\\

iii) subtraction of a hourly-minimum background from a raw white-light coronagraph image. This is also suitable for dynamic features. The background is built by taking the minimum value in each pixel of all images over a time range of 8 hours, centred at the launch time of the CME. As a result, we should obtain the brightness of the CME alone. Nevertheless, static features and the E-corona signal may still be present in the final image.\\

Within the field of view of COR1 the dominant signal is the K-corona (when compared with the F-corona). But because of the COR1 design (no external occulter) the instrumental stray-light is very high.

For COR2 (and LASCO) the F-corona is the dominant signal in the images. The instrumental stray-light in COR2 is very low (especially when compared to COR1).

As a consequence, the COR1 and COR2 monthly-minimum background images are fundamentally different. The COR1 monthly-minimum background image (subtracted as part of \textsf{secchi\_prep}) is used to remove instrumental stray-light. The COR2 monthly-minimum background images (not removed within \textsf{secchi\_prep}) are used to remove the F-corona signal.

In our study we used different background images for COR1 data. 
The COR1 monthly-minimum background subtracted images should leave us only with the K-corona signal. If any F-corona signal is left it should be unpolarised in the COR1 FOV and it should not influence the polarisation angle. Nevertheless, it is not guarantee that the full stray-light is removed by this background.
On the other side, when using the base-difference images, due to the short time between the two images, any contribution from stray-light and F-corona should be removed. But also part of the stable K-corona is removed, leaving us in theory only with the CME K-corona signal.
The hourly-minimum background subtraction, obtained over a time range of 8 hours centred at the launch time of the CME should remove all the unwanted signal and leave us only with the brightness of the CME alone, similar with base-difference images.

\section{The Influence of Different Background Subtractions} \label{S-bkginfl}

\begin{figure}
    \centering
	\includegraphics[width=0.9\textwidth]{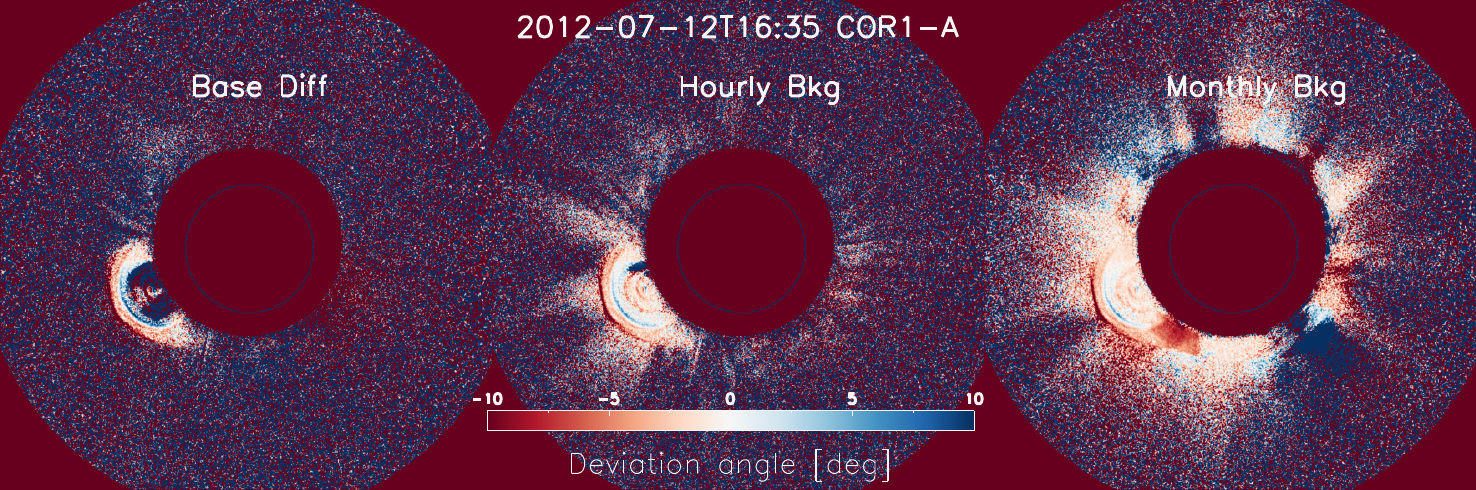}
	\caption{Deviation angle for COR1-A image at 16:35 with the base image subtracted (\textit{left panel}), hourly-minimum background subtracted (\textit{middle panel}) and monthly-minimum background subtracted (\textit{right panel}). }
	\label{F-appendix}
\end{figure}

Here we compare the COR1 base-difference images with the monthly-minimum background subtracted images and with the hourly-minimum background subtracted images (see Figure~\ref{F-appendix}), in order to see if different backgrounds influence our results.

The hourly-minimum background subtracted images are very similar with base-difference images, while the monthly-minimum background subtracted images show a poorer quality of the data. We suspect it is because the stray-light may be still present in these images.

As described in Appendix~\ref{S-bkg} no one of the above mentioned backgrounds is perfect. Nevertheless, all of them points to the same result: deviation higher than $10^\circ$ of the polarisation angle from tangential, in all pixels where a significant CME excess brightness is observed (see Figure~\ref{F-appendix}).

\bibliographystyle{spr-mp-sola}
\bibliography{bibliography}  

\end{article} 

\end{document}